\documentclass[aps,prd,preprint,groupedaddress,longbibliography,nofootinbib,showkeys]{revtex4-1}

\usepackage{amsmath}
\usepackage{graphicx}

\newcommand{\nt}{\notag}
\newcommand{\ep}{\epsilon}
\newcommand{\vev}[1]{\left\langle #1 \right\rangle}
\newcommand{\lb}{\left\lbrace}
\newcommand{\rb}{\right\rbrace}
\newcommand{\ket}[1]{\left| #1 \right\rangle}

\begin{document}

\preprint{KOBE-TH-22-07}

\title{Renormalization for a model of photon scattering \\off a charged
  harmonic oscillator}


\author{Hidenori~Sonoda}
\email[]{hsonoda@kobe-u.ac.jp}

\affiliation{Physics Department, Kobe University, Kobe 657-8501, Japan}


\date{27 December 2022}

\begin{abstract}
  In the electric dipole approximation the model of a charged harmonic
  oscillator interacting with the radiation field becomes quadratic
  and soluble, but it needs a UV cutoff for the photon frequency.  The
  model's renormalizability is only apparent; physics requires that
  the cutoff be kept finite.  The cutoff plays the role of a parameter
  that characterizes the high frequency behavior of the photon cross section.
\end{abstract}


\maketitle


\section{Brief introduction \label{introduction}}

Exponential decays of excited states were first derived from quantum
mechanics by Weisskopf and Wigner
\cite{Weisskopf:1930au,Weisskopf:1930ps}.  Their derivation makes it
clear that the decay has nothing to do with the non-linearity of the
system, but it is due to the mixing of discrete excited states with a
continuum of states involving photons.  This type of mixing has been
known as configuration interactions in the literature.

U.~Fano was the first to diagonalize a generic hamiltonian with a
configuration interaction \cite{Fano1961}\footnote{This is an
  extension of his earlier work \cite{Fano1935} whose English
  translation is available: U. Fano, G. Pupillo, A. Zannoni,
  C. W. Clark, J. Res. Natl. Inst. Stand. Technol. \textbf{110} (2005)
  583-587.}.  The solution is based on the method that Dirac used for
the derivation of resonance scattering \cite{Dirac:1927, Dirac:book}.  Fano
applied the result, known as Fano's profile, to the inelastic electron
scattering by He near the resonance corresponding to a double
excitation to $2s 2p$.

Fano's model has been studied as a simple model of renormalization in
\cite{Sonoda:2013dma}.  (Due to the author's ignorance,
\cite{Sonoda:2013dma} has no reference to Fano's earlier works.)  The
present paper is an extension, and differs technically from
\cite{Sonoda:2013dma} in two aspects.  First, the model is constructed
in the lagrangian formalism as opposed to the hamiltonian formalism.
Second, the time-ordered propagator of the oscillator coordinate is
considered as opposed to the retarded Green function of the
hamiltonian.  Otherwise, the same technique based on the analyticity
of the Green function is used.

The purpose of the paper is to gain further insights into the physics
of renormalization.  The outline of the paper is as follows.  In
Sec.~\ref{sec: the model} we introduce the model of photon scattering
off a charged harmonic oscillator.  We study the field theoretic
properties of the model as it is without questioning the validity of
the electric dipole approximation.  In Sec.~\ref{sec: sharp} we
introduce a sharp cutoff $\omega_H$ for the photon's frequency.  Two
renormalized parameters are defined: $\Omega_r$ and $r_{0,r}$.  We
show that $\omega_H$ has a maximum; hence, we cannot take the limit
$\omega_H \to +\infty$.  In Sec.~\ref{sec: scattering} we first sketch
the derivation of the photon scattering cross section
$\sigma (\omega)$ in terms of the spectral function of the propagator.
We then show how $\omega_H$ determines the behavior of
$\sigma (\omega)$ for $\Omega_r \ll \omega \ll\omega_H$.  In
Sec.~\ref{sec: wrong} we discuss the naive continuum limit
$\omega_H \to +\infty$.  We explain that the propagator obtains a
tachyon with a negative probability.  In Sec.~\ref{sec: smooth} we
explain that $\sigma (\omega)$ does not depend on the details of the
cutoff.  We show how to define $\omega_H$ for a smooth cutoff.  We
conclude the paper in Sec.~\ref{conclusion}.  In Appendix we explain
that the tachyon in the naive continuum limit is a consequence of the
negative kinetic term in the lagrangian.

\section{The model \label{sec: the model}}

We consider the lagrangian of a charged harmonic oscillator interacting
with the radiation field given by\footnote{We adopt the convention
  $\hbar = c = 1$.}
\begin{equation}
  L = \frac{m}{2} \frac{d}{dt} \vec{x} \cdot \frac{d}{dt} \vec{x} -
  \frac{m \Omega^2}{2} \vec{x}\,^2 + e \frac{d}{dt} \vec{x} \cdot
  \vec{A} (t, \vec{x}) +
  \int d^3 x\, \frac{1}{2} \left( \partial_t \vec{A} \cdot \partial_t \vec{A}
+  \vec{A} \cdot \nabla^2 \vec{A} \right)\,,
\end{equation}
where the vector potential satisfies the Coulomb gauge condition
\begin{equation}
    \nabla \cdot \vec{A} = 0\,.
\end{equation}
The lagrangian becomes quadratic and thus soluble in the electric
dipole approximation:
\begin{equation}
  L = \frac{m}{2} \frac{d}{dt} \vec{x} \cdot \frac{d}{dt} \vec{x} -
  \frac{m \Omega^2}{2} \vec{x}\,^2 + e \frac{d}{dt} \vec{x} \cdot
  \vec{A} (t, \vec{0}) +
  \int d^3 x\, \frac{1}{2} \left( \partial_t \vec{A} \cdot \partial_t \vec{A}
+  \vec{A} \cdot \nabla^2 \vec{A} \right)\,,\label{dipole}
\end{equation}
where the dependence of the vector potential on $\vec{x}$ is ignored
in the interaction term.  The approximation is valid as long as the
wave length of a photon is long compared with the size of the
oscillator:
\begin{equation}
  \lambda = \frac{2\pi}{\omega} > \frac{1}{\sqrt{m \Omega}}\,.
  \label{validity}
\end{equation}
In the following we examine the field theoretic properties of the
model given by (\ref{dipole}), where even the photons violating
(\ref{validity}) are included.  We are especially interested in
the question of renormalizability.
  
We expand
\begin{equation}
    \vec{A} (t, \vec{x}) = \frac{1}{\sqrt{V}} \sum_{\vec{k}} 
    \sum_{\vec{\ep}} \vec{\ep} \, \varphi_{\vec{k}, \vec{\ep}} \,(t)
    e^{i \vec{k} \cdot \vec{x}}\,,
\end{equation}
where $V$ is the space volume, $\vec{\ep}$ is a unit polarization
vector orthogonal to $\vec{k}$, and $\varphi_{\vec{k}, \vec{\ep}}$
satisfies
$\varphi_{-\vec{k}, \vec{\ep}} = \varphi_{\vec{k}, \vec{\ep}}^*$\,.
We then obtain
\begin{equation}
   \int d^3 x\, \frac{1}{2} \left( \partial_t \vec{A} \cdot
    \partial_t \vec{A} + \vec{A} \cdot \nabla^2 \vec{A} \right)
  = \sum_{\vec{k}} \sum_{\vec{\ep}} \frac{1}{2} \left(
    \partial_t \varphi_{-\vec{k}, \vec{\ep}}\, \partial_t \varphi_{\vec{k},
    \vec{\ep}}  - k^2 \varphi_{-\vec{k},\vec{\ep}} \, \varphi_{\vec{k},
    \vec{\ep}} \right)\,.
\end{equation}
Defining two real fields
\begin{equation}
  \phi_n \equiv \frac{1}{\sqrt{2}} \left( \varphi_{\vec{k}, \vec{\ep}}
    + \varphi_{-\vec{k}, \vec{\ep}} \right),\quad
  \psi_n \equiv \frac{1}{\sqrt{2}} \frac{1}{i}  \left( \varphi_{\vec{k}, \vec{\ep}}
    - \varphi_{-\vec{k}, \vec{\ep}} \right)\,,
\end{equation}
for $n = (\vec{k}, \vec{\ep})$, we obtain
\begin{equation}
   \int d^3 x\, \frac{1}{2} \left( \partial_t \vec{A} \cdot
    \partial_t \vec{A} + \vec{A} \cdot \nabla^2 \vec{A} \right)
  = \sum_n \frac{1}{2}
  \left( (\partial_t \phi_n)^2 - \omega_n^2 \phi_n^2 \right) +
  \sum_n \frac{1}{2} \left(  (\partial_t \psi_n)^2 - \omega_n^2 \psi_n^2 \right)\,,
\end{equation}
where $\omega_n = k \equiv |\vec{k}|$, and the sum is over a half of the
$\vec{k}$ space and $\vec{\ep}$.  (If $\vec{k}$ is included in the
sum, then $-\vec{k}$ is not.)  We will drop $\psi$'s with even parity
from now on, since only $\phi$'s with odd parity interact with the
charged oscillator.  Denoting $\vec{X} = \sqrt{m}\,\vec{x}$, we obtain
\begin{equation}
  L
  = \frac{1}{2} \frac{d}{dt} \vec{X} \cdot \frac{d}{dt} \vec{X} -
  \Omega^2 \frac{1}{2} \vec{X}\,^2 + \sum_n \frac{1}{2} \left(
    (\partial_t \phi_n)^2 - \omega_n^2 \phi_n^2 \right)
  + \frac{e}{\sqrt{m}} \frac{d}{dt}
  \vec{X} (t) \cdot \sqrt{\frac{2}{V}} \sum_n \vec{\ep}_n \phi_n
  (t)\,.
  \label{lagrangian}
\end{equation}
This is the model we are going to solve.

In the frequency space the free propagators are given by
\begin{subequations}
\begin{align}
  \vev{X_i (\omega) X_j (- \omega)}_0
  &= \delta_{ij} \frac{i}{\omega^2 - \Omega^2 + i \ep}\,,\\
  \vev{\phi_n (\omega) \phi_n (-\omega)}_0
  &= \frac{i}{\omega^2 - \omega_n^2 + i \ep}\,,
\end{align}
\end{subequations}
where the product is time ordered.  Hence, to the second order in
perturbation, the propagator of $\vec{X}$ is obtained as
\begin{align}
  &  \vev{X_i (\omega) X_j (-\omega)}\nt\\
  &\simeq \frac{i}{\omega^2 - \Omega^2 + i \ep} \left( \delta_{ij} -
    \frac{2 e^2}{m V} \omega^2 \sum_n \ep_{n,i} \ep_{n,j} \frac{i}{\omega^2
    - \omega_n^2 + i \ep} \frac{i}{\omega^2 - \Omega^2 + i \ep}
    \right)\,.
\end{align}
Summing over two polarization vectors, we obtain
\begin{equation}
  \sum_{\vec{\ep}} \ep_i \ep_j = \delta_{ij} - \frac{k_i k_j}{k^2}\,.
\end{equation}
Averaging over the directions of $\vec{k}$, we can replace this by
\begin{equation}
  \sum_{\vec{\ep}} \ep_i \ep_j = \frac{2}{3} \delta_{ij}\,.
\end{equation}
Hence, we obtain
\begin{equation}
  \sum_n \ep_{n,i} \ep_{n,j} \frac{1}{\omega^2 - \omega_n^2 + i \ep}
  = \frac{2}{3} \delta_{ij} \frac{1}{2} \sum_{\vec{k}} \frac{1}{\omega^2 - k^2 +
    i \ep}\,,
\end{equation}
where the factor $\frac{1}{2}$ is necessary since the sum over $n$ includes
only a half of the $\vec{k}$ space.  We then obtain
\begin{align}
  &  \vev{X_i (\omega) X_j (-\omega)}\nt\\
  &\simeq \delta_{ij} \frac{i}{\omega^2 - \Omega^2 + i \ep} \left( 1
  +  \frac{2 e^2}{3 m V} \omega^2 \sum_{\vec{k}}  \frac{1}{\omega^2
    - k^2 + i \ep} \frac{1}{\omega^2 - \Omega^2 + i \ep}
    \right)\,.
\end{align}

It is straightforward to go beyond the second order perturbation.
Summing the corresponding geometric series, we obtain the exact full
propagator as
\begin{equation}
\delta_{ij}\,  G (\omega^2 + i \ep) \equiv \frac{1}{i} \vev{X_i (\omega) X_j (-\omega)}
  = \delta_{ij} \frac{1}{\omega^2 - \Omega^2 - \frac{2 e^2}{3 mV}
    \omega^2 \sum_{\vec{k}} \frac{1}{\omega^2 - k^2 + i \ep}}\,.
\end{equation}
We now compute
\begin{align}
  \frac{2 e^2}{3 m V} \sum_{\vec{k}} \delta (k - \omega)
  &= \frac{2 e^2}{3 m V} V \int \frac{d^3 k}{(2 \pi)^3} \delta (k -
    \omega)\nt\\
  &= \frac{2 e^2}{3 m} \frac{4 \pi}{(2 \pi)^3} \omega^2 =   \frac{4}{3
    \pi} r_0 \omega^2\,, 
\end{align}
where
\begin{equation}
  r_0 \equiv \frac{e^2}{4 \pi} \frac{1}{m} \label{rzero}
\end{equation}
would be the classical electron radius if $m$ were the electron mass, and $e$ 
the elementary charge.  We can write the inverse propagator as
\begin{equation}
  \frac{1}{G (\omega^2 + i \ep)}
  = \omega^2 -
    \Omega^2 - \omega^2 \frac{4}{3 \pi} r_0
    \int_0^\infty d\omega'\,  \frac{\omega'\,^2 }{\omega^2 -
      \omega'\,^2 + i \ep} \,.\label{Ginv}
\end{equation}

\section{Sharp cutoff \label{sec: sharp}}

Now, (\ref{Ginv}) contains a UV divergent integral
\[
  \int_0^\infty d\omega'\, \frac{\omega'\,^2}{\omega^2-\omega'\,^2 + i
    \ep}\,.
\]
To make sense out of it, we need to introduce a high frequency cutoff
$\omega_H$:
\begin{align}
  \frac{1}{G (\omega^2 + i \ep)}
  &= \omega^2 - \Omega^2 - \omega^2
  \frac{4}{3 \pi} r_0 \int_0^{\omega_H} d\omega'
    \frac{\omega'\,^2}{\omega^2 - \omega'\,^2 + i \ep}\nt\\
  &= \omega^2 - \Omega^2 + \omega^2
    \frac{4}{3 \pi} r_0 \omega_H - \omega^4 \frac{4}{3 \pi} r_0
    \int_0^{\omega_H} d\omega'\, \frac{1}{\omega^2 - \omega'\,^2 + i
    \ep}\nt\\
  &= \left(1 + \frac{4}{3 \pi} r_0 \omega_H \right) \omega^2 -
    \Omega^2 - \omega^4 \frac{4}{3 \pi} r_0
    \int_0^{\omega_H} d\omega'\, \frac{1}{\omega^2 - \omega'\,^2 + i
    \ep}\,.
\end{align}
Initially, the lagrangian (\ref{lagrangian}) has only two parameters,
$\Omega$ and $r_0$, but the UV finiteness of the propagator requires
yet another parameter $\omega_H$.  We call the model renormalizable if
we can take the limit $\omega_H \to +\infty$.  For the time being we
only assume
\begin{equation}
  \Omega \ll \omega_H\,.
\end{equation}

For $0 < \omega < \omega_H$, we obtain
\begin{align}
   \int_0^{\omega_H} d\omega'\, \frac{1}{\omega^2 - \omega'\,^2 + i
    \ep}
  &= \frac{1}{2 \omega} \int_0^{\omega_H} d\omega'\, \left( \frac{1}{\omega - \omega' + i
    \ep} + \frac{1}{\omega+ \omega' - i \ep} \right)\nt\\
  &= \frac{1}{2 \omega} \int_{- \omega_H}^{\omega_H} d\omega'\,
    \frac{1}{\omega - \omega' + i \ep}\nt\\
  &= \frac{1}{2 \omega} \left( - \ln \frac{\omega_H - \omega}{\omega_H
    + \omega} - i \pi \right)\\
  &\overset{\omega \ll \omega_H}{\longrightarrow} - i \frac{\pi}{2
    \omega} + \frac{1}{\omega_H} + \mathrm{O}
    \left(\frac{\omega}{\omega_H^2}\right) \,.
\end{align}
Hence, for $0 < \omega \ll \omega_H$, we obtain
\begin{align}
  \frac{1}{G (\omega^2 + i \ep)}
  &\simeq \left(1 + \frac{4}{3 \pi} r_0
      \omega_H \right) \omega^2 - \Omega^2 - \frac{\omega^4}{\omega_H}
    \frac{4}{3\pi} r_0 + i \frac{2}{3} r_0 \omega^3\nt\\
  &= \left(1 + \frac{4}{3 \pi} r_0
    \omega_H \right) \left(
    \omega^2 - \Omega_r^2
    + r_{0, r} \omega^3
    \left(  \frac{2}{3} i - \frac{4}{3\pi} \frac{\omega}{\omega_H} \right)
\right)\,,
\end{align}
where we have defined two physical parameters by
\begin{align}
  \Omega_r^2
  &\equiv \frac{\Omega^2}{1 + \frac{4}{3 \pi} r_0 \omega_H}\,,\label{Omega-r}\\
  r_{0, r}
  &\equiv \frac{r_0}{1 + \frac{4}{3 \pi} r_0 \omega_H}\,.\label{rzero-r}
\end{align}
$\Omega_r$ is the approximate resonance frequency, and its full width
is given approximately by
\begin{equation}
  \Gamma = \frac{2}{3} r_{0,r} \Omega_r^2\,.
\end{equation}
We take
\begin{equation}
  r_{0, r} \Omega_r \ll 1
\end{equation}
so that the oscillator has a narrow width.\footnote{In the case of the
  photon scattering off a hydrogen atom, let the electron mass be $m$
  and the fine structure constant be $\alpha \equiv \frac{e^2}{4\pi}$.
  Then, $\Omega_r$ is of order $\alpha m$, and $r_{0,r}$ is the
  classical electron radius $\frac{\alpha}{m}$.  Hence,
  $r_{0,r} \Omega_r \sim \alpha^2 \ll 1$.  In this case $\omega_H \sim
m$ would be a good choice for the validity of the non-relativistic approximation.}

In addition to $\Omega_r$ and $r_{0,r}$, we introduce a wave function
renormalization for the harmonic oscillator
\begin{equation}
  \vec{X}_r \equiv \sqrt{1 + \frac{4}{3 \pi} r_0 \omega_H}\, \vec{X}
\end{equation}
so that the propagator is given by
\begin{equation}
  G_r (\omega^2 + i \ep) = \left(1 + \frac{4}{3 \pi} r_0
    \omega_H\right) G (\omega^2 + i \ep)\,.
\end{equation}
  
For $\omega > 0$ in general, we obtain
\begin{equation}
  G_r (\omega^2 + i \ep) = \lb
  \begin{array}{l@{\quad}l}
    \frac{1}{\omega^2 -
    \Omega_r^2 + \frac{1}{2} a \frac{\omega^3}{\omega_H} \left(\ln \frac{\omega_H
    - \omega}{\omega_H + \omega} + i \pi \right)}& ( 0 < \omega <
                                                   \omega_H )\,,\\
     \frac{1}{ \omega^2 -
    \Omega_r^2 -  \frac{1}{2} a \frac{\omega^3}{\omega_H}  \ln \frac{\omega
    +\omega_H}{\omega - \omega_H}} & ( \omega_H < \omega)\,,\\
    \frac{1}{\omega^2} \frac{1}{1 - a}  & (\omega \to \infty)\,,
  \end{array}\right.
\label{Grenormalized}
\end{equation}
where we have introduced a dimensionless parameter by
\begin{equation}
  a \equiv \frac{4}{3 \pi} r_{0,r} \omega_H = \frac{\frac{4}{3 \pi}
    r_0 \omega_H}{1 + \frac{4}{3 \pi} r_0 \omega_H}\,.\label{a-def}
\end{equation}
By definition, we find
\begin{equation}
  0 < a < 1\,.\label{a-range}
\end{equation}
Hence, given $r_{0,r}$, the highest cutoff we can take is finite:
\begin{equation}
  (\omega_H)_{\textrm{max}} = \frac{3\pi}{4 r_{0, r}}\,.\label{omegaH-max}
\end{equation}
This is an important result: we cannot take $\omega_H$ to infinity,
i.e., the model is not renormalizable.  As we discuss in
Sec.~\ref{sec: wrong} we can still force our way to take
$\omega_H \to \infty$, but in return we will end up with a model
plagued by a tachyon.

There are two artifacts due to the sharp edge of the cutoff:
\begin{enumerate}
\item \underline{a peak just below $\omega_H$} \\
  Let $\omega = \omega_H (1 - \eta)$, where $\eta$ is a small positive
  number.  We obtain, for small $a$,
\begin{equation}
  \frac{1}{G_r (\omega^2+i\ep)} \simeq \omega_H^2 \left[ 1 + 
\frac{a}{2}  \left( \ln \frac{\eta}{2} + i \pi \right) \right]\,.
\end{equation}
The real part vanishes at
\begin{equation}
  \eta = 2 e^{- \frac{2}{a}}\,,
\end{equation}
which is extremely small for small $a$.  Since $G_r$ vanishes at
$\omega = \omega_H$, the width of the second peak is of order
$\eta\, \omega_H = 2 e^{- \frac{2}{a}} \,\omega_H$.
\item \underline{a pole $\Omega_b$ just above $\omega_H$}\\
  Let $\Omega_b = \omega_H (1 + \eta)$.  We obtain, for small $a$,
\begin{equation}
  \frac{1}{G_r (\Omega_b^2)} \simeq \omega_H^2 \left[ 1 - \frac{a}{2} \ln
    \frac{2}{\eta} \right]\,.
\end{equation}
This vanishes at
\begin{equation}
 \eta = 2 e^{- \frac{2}{a}}\,.
\end{equation}
The pole is away from $\omega_H$ at the same distance as
the second peak on the other side.  The residue $r_b$ of the pole is
extremely small for small $a$:
\begin{equation}
  r_b \simeq \frac{8}{a} e^{- \frac{2}{a}} \ll 1\,.
  \end{equation}
\end{enumerate}
These two artifacts simply tell us that the sharp cutoff is only good
for low energy physics at $\omega \ll \omega_H$.

Since there is no pole on the negative real axis\footnote{There is no
  pole on the negative real axis; for $\omega^2 = - k^2 < 0$, we find
\[
    1/G_r (-k^2) = - k^2 \left( 1 - a + \frac{4}{3\pi} r_{0,r} \int_0^{\omega_H}
      d\omega' \frac{\omega'\,^2}{k^2+\omega'\,^2} \right)- \Omega_r^2 < 0\,,
  \]
  since $1-a>0$. }, we obtain the spectral representation
\begin{equation}
  G_r (\omega^2 + i \ep) = \int_0^{\omega_H} ds\, \rho (s) \frac{1}{\omega^2
    - s^2 + i \ep} + \frac{r_b}{\omega^2 - \Omega_b^2}\,,
\end{equation}
where
\begin{align}
  \rho (s)
  &= \frac{2 s}{\pi} (-) \Im G_r (s^2 + i \ep)\nt\\
  &= \frac{a \frac{s^4}{\omega_H}}{\lb  s^2 - \Omega_r^2 + a
    \frac{s^3}{\omega_H} \frac{1}{2}  \ln \frac{\omega_H-s}{\omega_H+s} \rb^2 +
    \left(\pi a  \frac{1}{2 \omega_H} s^3\right)^2}\\
  &\simeq \frac{4 r_{0,r}}{3 \pi} \frac{s^4}{(s^2-\Omega_r^2)^2 +
    \left( \frac{s}{\omega_H}\right)^2 \lb - 2 a s^2 (s^2-\Omega_r^2) +
    \left(\frac{\pi a}{2}\right)^2 s^4 \rb}\quad (s \ll \omega_H)\,.\label{rho-approx}
\end{align}
The asymptotic behavior (\ref{Grenormalized}) of $G_r (\omega^2+i\ep)$
gives the sum rule
\begin{equation}
  \int_0^{\omega_H} d\omega\, \rho (\omega) + r_b = \frac{1}{1 - a} >
  1\,.
  \label{rho-sum-rule}
\end{equation}
This is proportional to the sum of decay probabilities of the state
$\vec{X} \ket{\mathrm{vacuum}}$.  It decays either to a photon of
energy $0 < \omega < \omega_H$ or to the resonance at energy
$\Omega_b$.  The partial decay probabilities sum to $1$:
\begin{equation}
   \int_0^{\omega_H} d\omega\, (1-a) \,\rho (\omega) + (1-a) r_b = 1\,.\label{prob-sum}
\end{equation}

\section{Photon scattering with a sharp cutoff\label{sec: scattering}}

The spectral function $\rho$ not only gives the decay probability but
also gives the photon cross section.  For
completeness, we first sketch the derivation.  (See a
standard textbook such as \cite{sakurai1967}.)  We then discuss the
frequency dependence of the cross section.

\begin{figure}[h]
  \centering
  \includegraphics[width=0.45\textwidth]{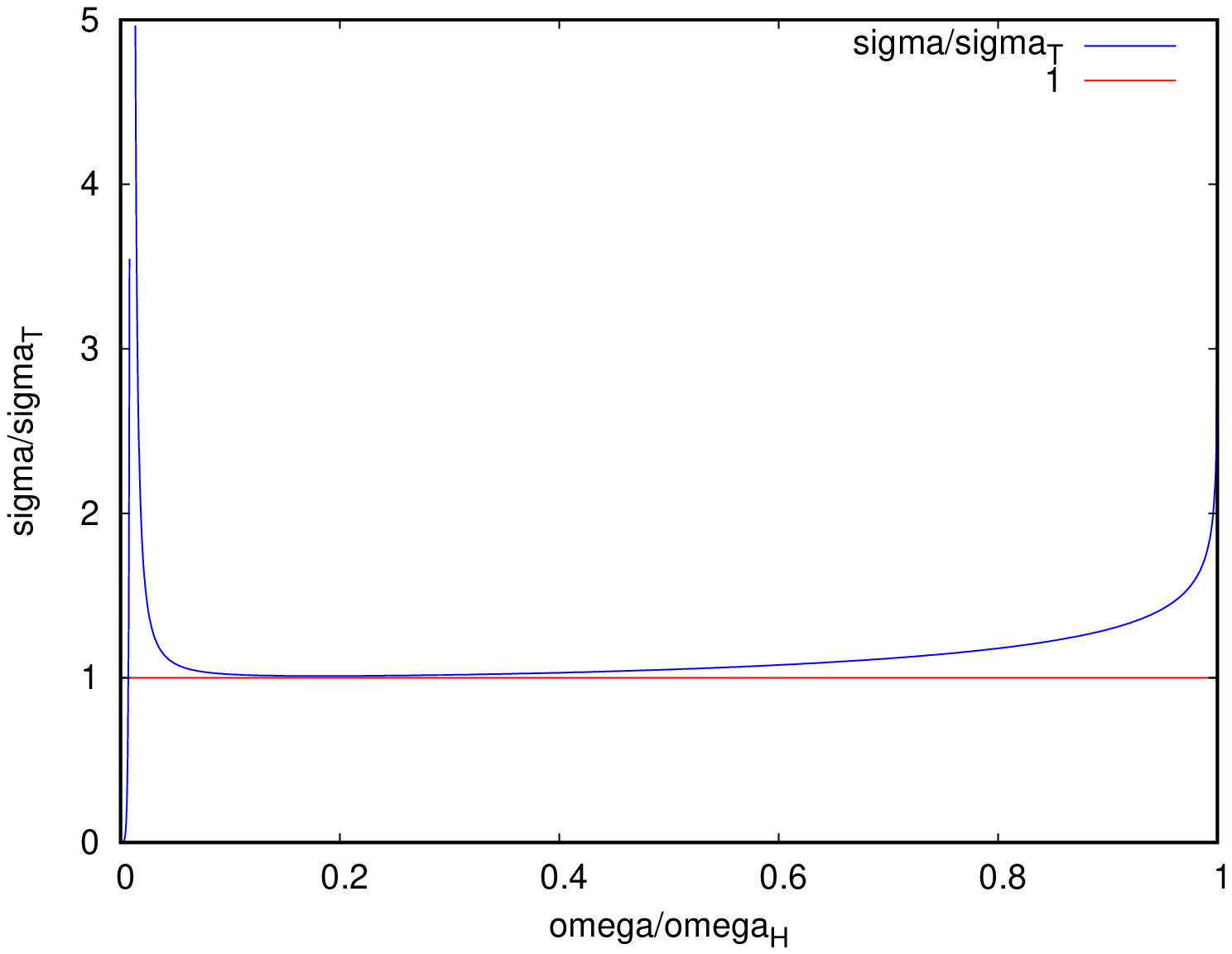}
    \includegraphics[width=0.45\textwidth]{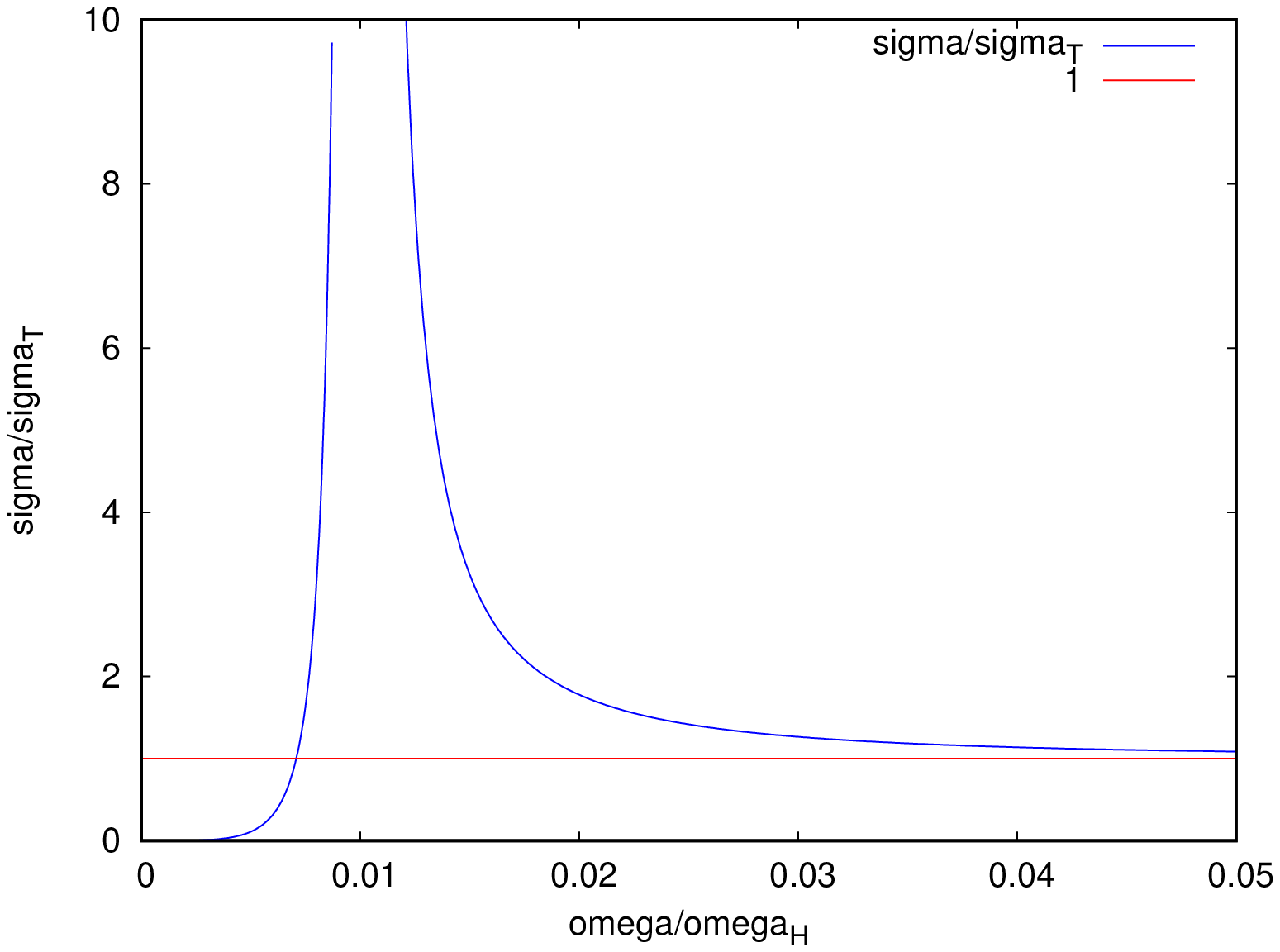}
  \caption{$\sigma (\omega)$ for $a \equiv \frac{4}{3 \pi} r_{0,r} \omega_H = 0.1$
    and $b \equiv \frac{\Omega_r}{\omega_H} = 0.01$.  $\sigma (\omega)$ vanishes at
    $\omega=\omega_H$; the peak just below $\omega=\omega_H$ is an
    artifact of the sharp cutoff.  The figure on the right is magnified
    for small $\omega$}
  \label{plot-sigma}
\end{figure}

To consider the photon scattering off the charged harmonic oscillator,
we compute
\begin{align}
  &  \vev{\phi_{\vec{k}', \vec{\ep}\,'} (\omega) \phi_{\vec{k}, \vec{\ep}} (- \omega)}\nt\\
  &= \delta_{\vec{k}, \vec{k}\,'} \delta_{\vec{\ep}, \vec{\ep}\,'}
    \frac{i}{\omega^2 - k^2 + i \ep} -
    \frac{e^2 \omega^2}{m V} \vec{\ep}\,' \cdot \vec{\ep} 
    \frac{i}{\omega^2 - k'\,^2 + i \ep} i G (\omega^2 + i \ep) 
    \frac{i}{\omega^2 - k^2 + i \ep}\nt\\
  &= \delta_{\vec{k}, \vec{k}\,'} \delta_{\vec{\ep}, \vec{\ep}\,'}
    \frac{i}{\omega^2 - k^2 + i \ep} - 4 \pi r_{0, r}
    \frac{\omega^2}{V} \vec{\ep}\,' \cdot \vec{\ep} 
    \frac{i}{\omega^2 - k'\,^2 + i \ep} i G_r (\omega^2 + i \ep) 
    \frac{i}{\omega^2 - k^2 + i \ep}\,.
\end{align}
This gives the transition matrix
\begin{equation}
  i T_{\vec{k}\,', \vec{\ep}\,'; \vec{k}, \vec{\ep}}
  = 2 \pi \delta (k - k') \frac{1}{\sqrt{2 k \cdot 2 k'}} (4 \pi
  r_{0,r}) \frac{k^2}{V} \left(\vec{\ep}\,' \cdot \vec{\ep}\right) (-  i) G_r (k^2
    + i \ep)
\end{equation}
and the transition probability per unit time
\begin{equation}
  w_{\vec{k}\,', \vec{\ep}\,'; \vec{k}, \vec{\ep}}
  =  2 \pi \delta (k - k') \frac{1}{4 k^2} (4 \pi r_{0,r})^2 \left(
    \frac{k^2}{V}\right)^2 \left(\vec{\ep}\,' \cdot \vec{\ep}\right)^2
  | G_r (k^2 + i \ep)|^2\,.
\end{equation}
Summing over the final states, we obtain
\begin{align}
  \sum_{\vec{k}\,', \vec{\ep}\,'}  w_{\vec{k}\,', \vec{\ep}\,'; \vec{k}, \vec{\ep}}
  &=  2 \pi \sum_{\vec{k}\,', \vec{\ep}\,'} \delta (k - k') \frac{1}{4
    k^2} (4 \pi r_{0,r})^2 \left( \frac{k^2}{V}\right)^2 \left(\vec{\ep}\,' \cdot
    \vec{\ep}\right)^2  | G_r (k^2 + i \ep)|^2\nt\\ 
  &= 2 \pi V \int \frac{d^3 k'}{(2 \pi)^3} \delta (k-k')
    \sum_{\vec{\ep}\,'} \frac{k^2}{4} \left(4 \pi r_{0,r}\right)^2 \frac{1}{V^2}
    \left(\vec{\ep}\,' \cdot \vec{\ep}\right)^2 |G_r (k^2 + i \ep)|^2\nt\\
  &= \frac{1}{V} \frac{4 \pi}{(2 \pi)^2} \frac{k^4}{4} (4 \pi r_{0,r})^2
    \underbrace{\sum_{\vec{\ep}\,'} \left(\vec{\ep}\,' \cdot \vec{\ep}\right)^2}_{\frac{2}{3}}
    |G_r (k^2 + i \ep)|^2\nt\\
  &= \frac{1}{V} \frac{8 \pi}{3} r_{0,r}^2 k^4 |G_r (k^2 + i \ep)|^2\,.
\end{align}
Since the flux is $\frac{1}{V}$, we obtain the desired result for the cross section as
\begin{align}
  \sigma (\omega)
  &= \frac{8 \pi}{3} r_{0,r}^2 \omega^4 |G_r (\omega^2 + i \ep)|^2\nt\\
  &= \frac{4 \pi}{\omega}  r_{0,r} \omega^2 (-) \Im G_r (\omega^2 + i \ep) =
    2 \pi^2 r_{0,r} \rho (\omega)\,.
\end{align}
This is valid for $\omega < \omega_H$.  See Fig.~\ref{plot-sigma}.
(\ref{rho-sum-rule}) gives the sum rule
\begin{equation}
  \int_0^{\omega_H} d\omega\, \sigma (\omega) = 2 \pi^2 r_{0,r}
  \left(\frac{1}{1 - a} - r_b\right) \simeq 2 \pi^2 \frac{r_{0,r}}{1 - a}\,.\label{sigma-sum-rule}
\end{equation}

For $\omega \ll \omega_H$, (\ref{rho-approx}) gives
\begin{equation}
  \sigma (\omega) \simeq \sigma_T
  \frac{\omega^4}{\left(\omega^2-\Omega_r^2\right)^2 +
    \left(\frac{2 r_{0,r}}{3}\right)^2 \omega^4 \lb
   \omega^2 - \frac{8}{\pi^2 a} (\omega^2-\Omega_r^2)\rb}\,,
\end{equation}
where $\sigma_T$ is the Thomson cross section:
\begin{equation}
  \sigma_T \equiv \frac{8\pi}{3} r_{0,r}^2\,.
\end{equation}
As $\omega$ grows beyond $\Omega_r$, $\sigma (\omega)$ approaches $\sigma_T$ (right
of Fig.~\ref{plot-sigma}) until the second term in the denominator
starts giving a growing negative contribution.  $\sigma (\omega)$
rises slowly as $\omega$ grows toward $\omega_H$ (left of
Fig.~\ref{plot-sigma})\footnote{For $\frac{8}{\pi^2} < a \equiv \frac{4}{3 \pi} r_{0,r}
  \omega_H < 1$, $\sigma (\omega)$ is decreasing in this range.}:
\begin{equation}
  \sigma (\omega) \simeq \sigma_T \left( 1 + \omega^2 \left(\frac{2
        r_{0,r}}{3}\right)^2 \left( \frac{6}{\pi} \frac{1}{r_{0,r} \omega_H} - 1 \right)\right)
  \quad
    \left( \Omega_r \ll \omega \ll \omega_H\right) \,.\label{sigma-omegaH}
  \end{equation}
  The presence of the frequency cutoff $\omega_H$ affects the $\omega$
  dependence of the cross section.

\section{What is wrong with the naive continuum limit\label{sec: wrong}}

To recapitulate, we have shown that the renormalized propagator
$G_r (\omega^2 + i \ep)$ depends on three parameters: $\Omega_r$
giving the resonance frequency, $r_{0,r}$ giving the Thomson cross
section, and $\omega_H$ giving the high frequency cutoff of the
photons.  The photon scattering cross section at high frequencies
depends on $\omega_H$ as given by (\ref{sigma-omegaH}).

Now, (\ref{omegaH-max}) tells us that the model is not renormalizable
since $\omega_H$ cannot be taken to infinity.  But if we examine the
renormalized propagator given by (\ref{Grenormalized}), we find
\begin{equation}
  G_r (\omega^2 + i \ep)
  = \frac{1}{\omega^2 - \Omega_r^2 + \frac{2}{3 \pi} r_{0,r} \omega^3
    \left(\ln \frac{\omega_H-\omega}{\omega_H+\omega} + i \pi \right)}
\end{equation}
has a naive limit
\begin{equation}
  \lim_{\omega_H \to +\infty}
  G_r (\omega^2 + i \ep)
  = \frac{1}{\omega^2 - \Omega_r^2 + \omega^3 \frac{i \pi}{2}
    \frac{4}{3 \pi} r_{0, r}}\,.
\end{equation}
What is wrong with this limit?

Let us examine the limit on the negative axis of $\omega^2$.  At
$\omega^2 = - k^2 < 0$ we find
\begin{align}
  \frac{1}{G_r (-k^2)}
  &= - k^2 - \Omega_r^2 + k^4 \frac{4}{3 \pi} r_{0,r}
    \int_0^\infty d\omega'\, \frac{1}{k^2 + \omega'\,^2}\nt\\
  &= - k^2 - \Omega_r^2 + k^3 \frac{2}{3} r_{0,r}\,.\label{negative-axis}
\end{align}
This has a zero at
$k = k_t \simeq \frac{3}{2 r_{0,r}} + \frac{2}{3} r_{0,r} \Omega_r^2>
0$, where
\begin{equation}
  \frac{d}{dk} \frac{1}{G (-k^2)} \Big|_{k=k_t} = \frac{2 k_t}{r_t} > 0\,.
\end{equation}
(The positive constant $r_t$ is determined shortly.)  This implies
\begin{equation}
  \frac{1}{G_r (-k^2)} \simeq \frac{2 k_t}{r_t} (k - k_t) \Longrightarrow
  G_r (-k^2) \simeq \frac{r_t}{2 k_t(k-k_t)} \simeq 
   \frac{- r_t}{-k^2 + k_t^2}\,.
\end{equation}
Hence, the propagator $G_r (\omega^2+i \ep)$ has a tachyon pole at
$\omega^2 = - k_t^2 < 0$, and the negative residue $- r_t$ at the pole
implies a negative probability.

\begin{figure}[h]
  \centering
  \includegraphics[width=0.5\textwidth]{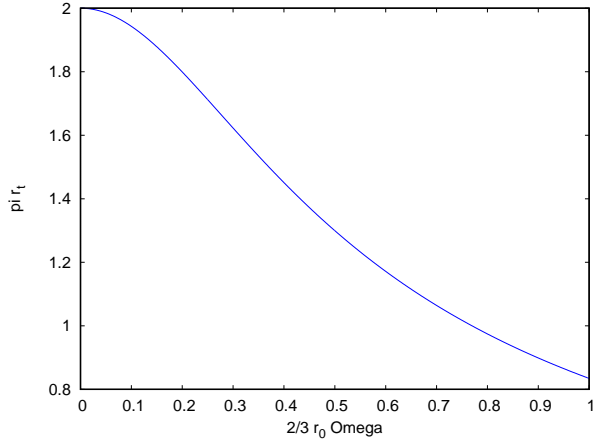}
  \caption{The residue of the tachyon pole is $- \pi r_t$.  $\pi r_t$ is converging
    toward $2$ as $\frac{2}{3} r_{0,r} \Omega_r \to 0+$, though
    $\pi r_t = 1$ strictly at zero}
  \label{fig-tachyon}
\end{figure}

The spectral representation becomes
\begin{equation}
  G_r (\omega^2 + i \ep) = - \frac{r_t}{\omega^2 + k_t^2} +
  \int_0^\infty \frac{ds}{\pi} \frac{s^4 \frac{4}{3\pi} r_{0,r}}{\left(s^2
      - \Omega_r^2\right)^2 + s^6 \left(\frac{2 r_{0,r}}{3}\right)^2}
  \frac{1}{\omega^2 - s^2 + i \ep}\,.
\end{equation}
The sum rule gives the residue of the tachyon pole as
\begin{align}
  r_t
  &=  \int_0^\infty \frac{ds}{\pi} \frac{s^4 \frac{4}{3\pi} r_{0,r}}{\left(s^2
    - \Omega_r^2\right)^2 + s^6 \left(\frac{2 r_{0,r}}{3}\right)^2}\nt\\
  &= \frac{2}{\pi^2} \int_0^\infty dx\, \frac{x^4}{\left( x^2 -
    \left(\frac{2}{3} r_{0,r} \Omega_r\right)^2 \right)^2 + x^6}\,.
\end{align}
This is discontinuous at $\frac{2}{3} r_{0,r} \Omega_r = 0$; for small
$r_{0,r} \Omega_r \ll 1$, $r_t \simeq \frac{2}{\pi}$, but
$r_t = \frac{1}{\pi}$ at $r_{0,r} = 0$.  We plot $\pi r_t$ for
$0.001 < \frac{2}{3} r_{0,r} \Omega_r < 1$ in Fig.~\ref{fig-tachyon}.

Now, what is wrong with the tachyon?  The tachyon pole in
$G (\omega^2 + i \ep)$ implies the presence of a \textbf{negative
  norm} state with purely imaginary energy $\mp i k_t$.  This is a
clear violation of unitarity of the theory.  In Appendix we explain a
little more about the tachyon: we can trace its origin to a negative
sign of the kinetic term in the lagrangian.

\section{Smooth cutoff \label{sec: smooth}}

Instead of introducing a sharp cutoff $\omega_H$ we may introduce a
smooth positive cutoff function $K (\omega)$ that decays smoothly as
$\omega \to +\infty$.  As we will see, the physics at frequencies
sufficiently smaller than $\omega_H$ does not depend on which cutoff
scheme we use.  This is universality.

\begin{figure}[h]
  \centering
  \includegraphics{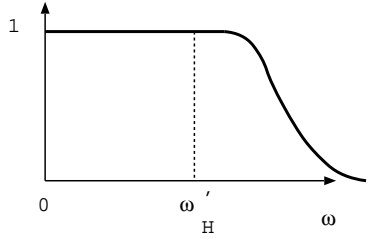}
  \caption{Cutoff function $K(\omega)$ is $1$ for low frequencies}
  \label{fig-cutoff}
\end{figure}

Eq.~(\ref{Ginv}) is replaced by
\begin{equation}
  \frac{1}{G (\omega^2 + i \ep)} = \omega^2 - \Omega^2 - \omega^2
  \frac{4}{3 \pi} r_0 \int_0^\infty d\omega'\, K (\omega')
  \frac{\omega'\,^2 }{\omega^2 - \omega'\,^2 + i \ep}\,,
\end{equation}
where $K (\omega) = 1$ for low frequencies, but it decays fast
enough as $\omega \to \infty$.  (See
Fig.~\ref{fig-cutoff}.)\footnote{Fast enough so that
  $\int_0^\infty d\omega\, \omega^2 K(\omega) < + \infty$.}
We obtain
\begin{equation}
  \frac{1}{G (\omega^2 + i \ep)}
  = \left(1 + \Delta z  \right) \omega^2 - \Omega^2 - \omega^4 \frac{4}{3\pi} r_0 \int_0^\infty
  d\omega'\, K(\omega') \frac{1}{\omega^2 - \omega'\,^2 + i \ep}\,,
\end{equation}
where we have defined
\begin{equation}
\Delta z \equiv  \frac{4 r_0}{3 \pi} \int_0^\infty d\omega' 
K(\omega') > 0 \,.\label{Delta-z}
\end{equation}
As in Sec.~\ref{sec: sharp} we introduce renormalized quantities as
follows:
\begin{subequations}
  \begin{align}
    \Omega_r^2
    &\equiv \frac{\Omega^2}{1 + \Delta z}\,,\\
    r_{0, r}
    &\equiv \frac{r_0}{1 +\Delta z}\,,\label{rzero-K}\\
    G_r (\omega^2+i\ep)
    &\equiv \left(1 + \Delta z \right) G (\omega^2 + i \ep)\,,
  \end{align}
\end{subequations}
so that
\begin{equation}
    \frac{1}{G_r (\omega^2 + i \ep)} = \omega^2 - \Omega_r^2 - \omega^4
  \frac{4}{3 \pi} r_{0,r} \int_0^\infty d\omega'\, K(\omega')
  \frac{1}{\omega^2 - \omega'\,^2 + i \ep}\,.\label{Gr-K}
\end{equation}
(\ref{Delta-z}) and (\ref{rzero-K}) imply that $a_K$ defined below
is less than $1$:
\begin{equation}
a_K \equiv  \frac{4}{3 \pi} r_{0,r} \int_0^\infty d\omega'\, K(\omega') =
  \frac{\Delta z}{1+\Delta z} < 1\,.  \label{inequality-K}
\end{equation}
Defining
\begin{equation}
  b_\omega \equiv \omega^2 - \Omega_r^2 - \omega^4 \frac{4}{3 \pi} r_{0,r}
  \int_0^\infty d\omega'\, K(\omega') \mathbf{P}
  \frac{1}{\omega^2 - \omega'\,^2}\label{bomega-K}
\end{equation}
for $\omega > 0$, we obtain
\begin{equation}
  G_r (\omega^2 + i \ep)
  = \frac{1}{b_\omega + \omega^4 \frac{4}{3 \pi} r_{0,r} K(\omega)
    \frac{i\pi}{2 \omega}}\,.\quad (\omega > 0)
\end{equation}

As in Sec.~\ref{sec: sharp} we can obtain a spectral representation of
$G_r$:
\begin{equation}
  G_r (\omega^2 + i \ep) = \int_0^\infty ds\, \rho (s)
  \frac{1}{\omega^2 - s + i \ep}\,,
\end{equation}
where the positive spectral function is given by
\begin{align}
  \rho (s)
  &= \frac{2 s}{\pi} (-) \Im G_r (s^2 + i \ep)\nt\\
  &= \frac{s^4 \frac{4}{3 \pi} r_{0,r} K(s)}{b_s^2 + \left(s^4
    \frac{4}{3\pi} r_{0,r} K(s) \frac{\pi}{2s}\right)^2}\,.
\end{align}

To obtain a sum rule for $\rho (s)$, we must make sure that
$G_r (\omega^2)$, defined by (\ref{Gr-K}) for the entire complex plane
of $\omega^2$, has no pole on the negative real axis.  There is no
such pole because at $\omega^2 = - k^2 < 0$ we obtain
\begin{equation}
  - k^2 \left( 1 -  \frac{4}{3 \pi} r_{0,r} \int_0^\infty d\omega'\,
    K(\omega') \frac{k^2}{k^2 + \omega'\,^2} \right) -
  \Omega_r^2 < 0
\end{equation}
thanks to (\ref{inequality-K}).
The asymptotic behavior
\begin{equation}
  G_r (\omega^2 + i \ep) \overset{\omega^2 \to \infty}{\longrightarrow}
  \frac{1}{1 - a_K} \frac{1}{\omega^2}
\end{equation}
then implies a sum rule analogous to (\ref{rho-sum-rule}):
\begin{equation}
  \int_0^\infty ds\, \rho (s) = \frac{1}{1 - a_K}\,.
\end{equation}
Accordingly, the cross section
$\sigma (\omega) = 2 \pi^2 r_{0,r} \rho (\omega)$ satisfies
\begin{equation}
  \int_0^\infty d\omega\, \sigma (\omega) = 2 \pi^2 r_{0,r} \frac{1}{1
    - a_K}\,,
  \label{sigma-sum-rule-K}
\end{equation}
analogous to (\ref{sigma-sum-rule}).

\begin{figure}[h]
  \centering
  \includegraphics[width=0.8\textwidth]{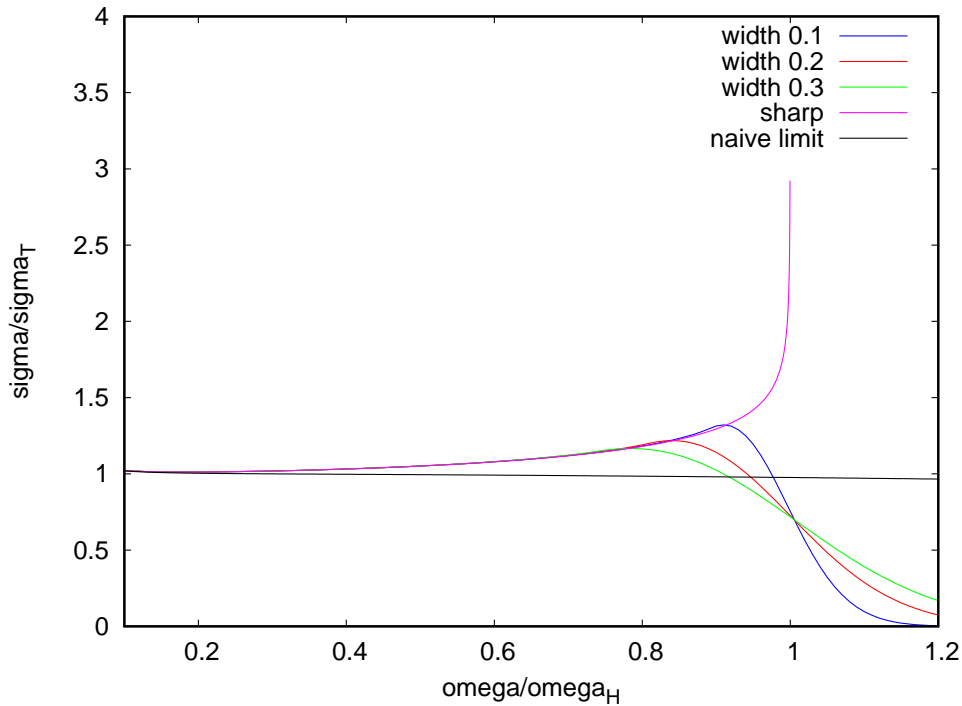}
  \caption{$\sigma (\omega)$ for $0.1 < \frac{\omega}{\omega_H} < 1.2$
    with four choices of cutoff.  The ``width'' of a smooth cutoff is
    the width of the region where $K$ drops from $1$ to almost $0$.
    The four schemes agree for $\frac{\omega}{\omega_H} < 0.8$.  We
    also show the ``naive limit'' discussed in Sec.~\ref{sec: wrong}}
  \label{fig-sigma}
\end{figure}

We would like to find the approximate behavior of the cross section
for $\omega$ small compared with a cutoff scale.  Suppose
$K (\omega) = 1$ for $\omega < \omega'_H$, where $\omega'_H$ is much
larger than $\Omega_r$.  We then obtain, for $0 < \omega < \omega'_H$,
\begin{align}
&  \int_0^\infty d\omega'\, K(\omega') \frac{1}{\omega^2 -
                 \omega'\,^2 + i \ep}\nt\\
  &= \int_0^{\omega'_H} d\omega'\,  \frac{1}{\omega^2 -
    \omega'\,^2 + i \ep} + \int_{\omega'_H}^\infty d\omega'\,
    K(\omega') \frac{1}{\omega^2 - \omega'\,^2}\nt\\
  &= \frac{1}{2\omega} \left( - \ln
    \frac{\omega'_H-\omega}{\omega'_H+\omega} - i \pi \right)
   + \int_{\omega'_H}^\infty d\omega'\,
    K(\omega') \frac{1}{\omega^2 - \omega'\,^2}\,.
\end{align}
Expanding in powers of $\frac{\omega}{\omega'_H}$, we obtain
\begin{equation}
  \int_0^\infty d\omega'\, K(\omega') \frac{1}{\omega^2 -
    \omega'\,^2 + i \ep}
  \simeq - \frac{i \pi}{2 \omega} + \frac{1}{\omega'_H} -
  \int_{\omega'_H}^\infty d\omega'\, K(\omega') \frac{1}{\omega'\,^2}\,.
\end{equation}
Hence, for $\omega \ll \omega'_H$, we can approximate
\begin{equation}
  b_\omega \simeq \omega^2 - \Omega_r^2 - \omega^4 \frac{4}{3 \pi} r_{0,r}
  \left( \frac{1}{\omega'_H} -  \int_{\omega'_H}^\infty d\omega'\,
    K(\omega') \frac{1}{\omega'\,^2} \right)\,.
\end{equation}
We can identify
\begin{equation}
\frac{1}{\omega_H} \equiv  \frac{1}{\omega'_H} -  \int_{\omega'_H}^\infty d\omega'\,
  K(\omega') \frac{1}{\omega'\,^2}\label{omegaH-K}
\end{equation}
with the cutoff frequency of the sharp cutoff.  As long as
$\omega \ll \omega_H$, the smooth cutoff gives the same propagator,
hence the same cross section, as the sharp cutoff with the same
$\omega_H$.  We note that $\omega_H$ given by (\ref{omegaH-K}) depends
only on the cutoff function $K$ but not on the particular choice of
$\omega'_H$.\footnote{Given $K$, $\omega'_H$ is not uniquely
  determined.  The largest possible $\omega'_H$ is determined by $K$.}
We show the cross section $\sigma (\omega)$ for three smooth cutoff
functions and a sharp cutoff corresponding to the same $\omega_H$.
(Fig.~\ref{fig-sigma}) They seem to agree even up to relatively large
$\omega$.

In conclusion, the cross section for $\omega$ sufficiently smaller than
$\omega_H$ is independent of a particular choice of the cutoff scheme.
The sum rule (\ref{sigma-sum-rule-K}) for the cross section, however,
depends on the cutoff scheme.

\section{Conclusion \label{conclusion}}

In this paper we have reconsidered the familiar model of a charged
harmonic oscillator interacting with the radiation field in the
electric dipole approximation, and examined the question of
renormalizability.  

The lagrangian (\ref{lagrangian}) with two parameters
$\Omega, r_0 = \frac{e^2}{4 \pi m}$ is naively renormalizable via the
renormalization of the wave function and the parameters.  But we have
identified the problem of a tachyon that spoils unitarity of the
model.  The frequency cutoff $\omega_H$ must be kept finite, and it
can be considered as the third parameter of the model besides the
renormalized parameters $\Omega_r$ and $r_{0,r}$, defined by
(\ref{Omega-r}, \ref{rzero-r}).  We have shown that for frequencies
sufficiently smaller than $\omega_H$, the photon scattering cross
section depends only on the three parameters, not sensitive to the
precise way the cutoff is introduced.  In this sense $\omega_H$ is
analogous to the coupling constant of many renormalizable theories in
four dimensions, such as $\phi^4$ and QED, whose renormalizability is
only perturbative.  The problem with tachyons is well known in the
renormalization of the Lee model \cite{Lee:1954iq} and the large $N$
limit of the $\phi^4$ theory \cite{Coleman:1974jh}, both in four
dimensions.  The tachyons are absent as long as the UV cutoff is kept
finite in both cases.

\appendix*

\section{The origin of a tachyon}

We would like to show that the tachyon we have found in the naive
continuum limit can be traced to a negative kinetic term in the
lagrangian.  In terms of renormalized parameters the lagrangian
(\ref{lagrangian}) of the model is given by
\begin{equation}
  L = \frac{1}{2} (1 - a) \frac{d}{dt} \vec{X}_r \cdot \frac{d}{dt}
  \vec{X}_r - \Omega_r^2 \frac{1}{2} \vec{X}_r^2 + \sqrt{4 \pi
    r_{0,r}}\, \frac{d}{dt} \vec{X}_r \cdot \sqrt{\frac{2}{V}} \sum_n
  \vec{\ep}_n \phi_n + \sum_n \frac{1}{2} \left( \left(\partial_t
      \phi_n\right)^2 - \omega_n^2 \phi_n^2 \right)\,,
\end{equation}
where the parameter $a$ is given by
\[
  a \equiv \frac{4}{3\pi} r_{0,r} \omega_H\,.\eqno{(\ref{a-def})}
\]
Given $r_{0,r}$, disregarding the inequality (\ref{a-range}), we
may increase $\omega_H$ beyond the limit (\ref{omegaH-max}).  For
$a > 1$, the kinetic term is negative, and we are not surprised to
find that the propagator acquires a tachyon with a negative residue.

To be more precise, on the negative axis of $\omega^2$, the inverse
of the renormalized propagator $G_r(\omega^2+i\ep)$ is given by
\begin{equation}
  \frac{1}{G_r (-k^2)} = - k^2 - \Omega_r^2 + k^3 \frac{4 r_{0,r}}{3 \pi}
     \arctan \frac{\omega_H}{k}\,.
\end{equation}
In the limit $\omega_H \to +\infty$, this gives (\ref{negative-axis})
considered in Sec.~\ref{sec: wrong}.  We can show that this propagator has
a simple pole at $k^2 = k_t^2$ with a negative residue $- \pi r_t$.
For $a$ slightly larger than $1$, we obtain approximately
\begin{equation}
  \frac{k_t}{\omega_H} \simeq \sqrt{\frac{\frac{a}{3} +
      \left(\frac{\Omega_r}{\omega_H}\right)^2}{a-1}}\,,\quad
  r_t \simeq \frac{1}{a-1}\,.
\end{equation}
We have discussed the limit $a \to +\infty$ in Sec.~\ref{sec: wrong}.  As
$a$ increases toward $1$, the bound state pole $\Omega_b^2$, discussed
in Sec.~\ref{sec: sharp}, approaches infinity.  As we take $a$ across $1$,
the pole returns as a tachyon pole $-k_t^2$, and it moves toward the
pole found in Sec.~\ref{sec: wrong} as we increase $a$ further.

\begin{acknowledgments}
  The undergraduate seminar series I conducted during the academic
  year 2021 gave me an opportunity to reconsider this familiar model
  of photon scattering.  I would like to thank my students (Y.~Arai,
  R.~Atsumi, K.~Kishimoto, K.~Lee) for their active participation in
  the seminars.
\end{acknowledgments}

\bibliography{paper} 

\end{document}